\documentclass[12pt,a4paper]{article}

\usepackage{amsfonts,amssymb,amsmath,amsopn,amsthm,graphicx,color,a4}

\title{Spectral Properties of Finite Quantum Hall Systems\thanks{Accepted for J. Oper. Theor.}}
\author{Christian Ferrari and Nicolas Macris}
\date{Institut de Physique Th\'eorique \\ Ecole Polytechnique F\'ed\'erale \\
CH - 1015 Lausanne, Switzerland}

\numberwithin{equation}{section}

\newtheorem{thm}{Theorem}
\newtheorem{hyp}{Hypothesis}

\DeclareMathOperator{\supp}{supp}
\DeclareMathOperator{\dist}{dist}

\begin{document}

\maketitle

\abstract{In this note we review spectral properties of
magnetic random Schr\"odinger operators $H_\omega=H_0+V_\omega + U_\ell + U_r$
defined on $L^2(\mathbb{R}\times\left[-\frac{L}{2},\frac{L}{2}\right],\,\textrm{d} x \,\textrm{d} y)$ with periodic boundary
conditions along $y$. $U_\ell$ and $U_r$ are two confining potentials 
for $x\leq -\frac{L}{2}$ and $x\geq \frac{L}{2}$ respectively and vanish for
$-\frac{L}{2}\leq x \leq \frac{L}{2}$. We describe the spectrum in two energy intervals and
we classify it according to the quantum mechanical current of eigenstates along the periodic direction.
The first interval lies in the first Landau band of the bulk Hamiltonian,
and contains intermixed eigenvalues with a quantum mechanical current of ${\cal O}(1)$ and ${\cal O}\left(e^{-\gamma B(\log
L)^2}\right)$ respectively. The second interval lies in the first spectral gap of the 
bulk Hamiltonian, and contains only eigenvalues with a quantum
mechanical current of ${\cal O}(1)$.}

\section{Introduction}
\indent
In this note we review recent results on the
spectrum of a magnetic random Schr\" odinger operator $H_\omega$ which describes the dynamics of an
electron lying on a cylinder of circumference $L$ and which is confined along
the cylinder axis by two smooth increasing potentials whose supports are
separated by a distance $L$. We suppose our particle spinless, thus
the Zeeman term in the Hamiltonian is neglected.
The complete proofs of the theorems stated here can be found  in \cite{FM1} and \cite{FM2}.

First let us shortly recall previous results on random
Schr\" odinger operators with magnetic field in the infinite two dimensional plane $\mathbb{R}^2$. We denote by
$H_0$ the kinetic term $H_0=(p-A)^2$, where $A$ is the vector
potential associated to a constant magnetic field $B$. The spectrum of $H_0$
is given by the Landau levels $\left\{(2n+1)B: n\in \mathbb{N}\right\}$.
The bulk Hamiltonian is
\begin{equation}\label{bulk-h}
H_\omega^b=H_0+V_\omega
\end{equation}
where $V_\omega$ is a Anderson-like random potential. The spectrum of
\eqref{bulk-h} is contained in Landau bands around each Landau level, 
$\sigma(H^b_\omega) \subset \bigcup_{n\geq 0} \left[(2n+1)B - V_0,(2n+1)B + V_0\right]$
where $V_0=\|V_\omega\|$, and if $V_0<B$ there are open spectral gaps
$G_n\supseteq\left((2n+1)B+V_0,(2n+3)B-V_0\right)$ ($n\in \mathbb{N}$).
It is proven that near the band edges the spectrum of $H_\omega^b$ is almost
surely pure point with exponentially localized eigenfunctions \cite{DMP1}, \cite{DMP2}, \cite{CH},
\cite{BCH}, \cite{W}. There are
no rigorous results for energies at the band centers, except for a special model
where the impurities are point scatterers \cite{DMP3}, \cite{DMP4}. 

We now add a wall potential, translation invariant along the $y-$direction, such
that $U_\ell(x)$ is confining for $x\leq -\frac{L}{2}$ and $U_\ell(x)=0$ for $x\geq
-\frac{L}{2}$. We have a semi-infinite system with a Hamiltonian 
\begin{equation}\label{1edge-h}
H_\omega^{si}=H_0+V_\omega + U_\ell \; .
\end{equation}
The spectrum contains the interval $[B,+\infty)$. For this system one can show that,
for energies in intervals inside the gaps of the bulk Hamiltonian, the average velocity $(\psi,v_y\psi)$
in the $y$ direction, of an assumed eigenstate $\psi$ does not
vanish. Since the velocity $v_y$ is the commutator between $-iy$ and the
Hamiltonian, the Virial Theorem implies that an eigenstate
cannot exist, and that therefore the
spectrum is purely continuous inside the gaps of the bulk Hamiltonian \cite{MMP}, \cite{F}. By Mourre
theory one can show that the spectrum therein is purely absolutely
continuous \cite{FGW}, \cite{dBP}. 

Finally we can add a second wall potential $U_r$ such that $U_r(x)=0$ for $x\leq
\frac{L}{2}$ and which is confining for $x\geq \frac{L}{2}$. So the particle is
confined between $x=-\frac{L}{2}$ and $x=\frac{L}{2}$. The Hamiltonian has the
form 
\begin{equation}
H_\omega=H_0+V_\omega + U_\ell + U_r \; .
\end{equation}

\begin{figure}[!h]\label{model}
\begin{center}
\input{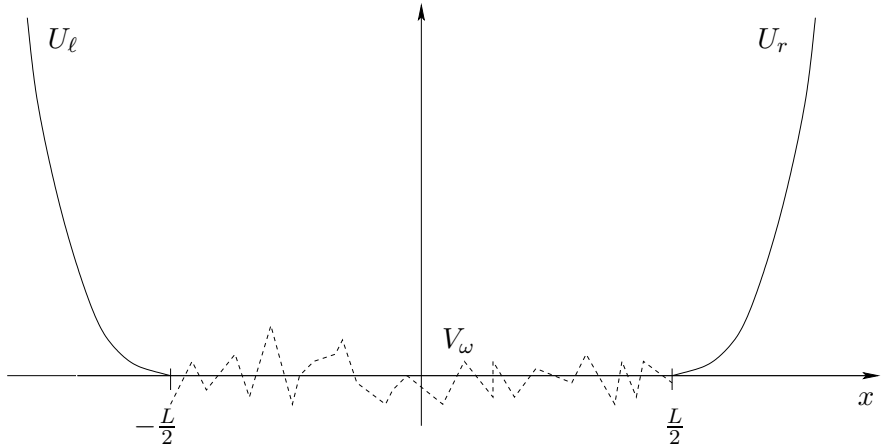}
\end{center}
\caption{\small{\emph{The potentials along the $x$-axes.}}}
\end{figure}

Additionally we make the $y-$direction periodic of length $L$ ($L$ a large
parameter), this correspond to a finite, macroscopic cylindrical geometry.

Other models with confinement in the $x-$direction but without p.b.c. along $y$ have been studied. The first consists of a parabolic
channel where $U_\ell+U_r$ is replaced by a parabolic confinement $\gamma x^2$,
in this case it is shown that if the perturbation ($V_\omega$ in our case) is
small enough in a suitable sense and satisfies a weak decay
condition in the $y-$direction, there exists intervals of absolutely continuous
spectrum \cite{EJK}. The second model, more close to ours, consists to take two step walls of finite
height for $U_\ell$ and $U_r$. In this case an initial state localized in
energy in the spectral gap of the bulk Hamiltonian and near the left (resp.
right) wall has a positive (resp. negative) velocity up to a finite time,
limited by tunneling effect between the two walls \cite{C}.

The physical interest of our model is related to the integral
quantum Hall effect \cite{PG}. For the
explanation of this effect Halperin \cite{H} pointed out the
importance of the boundary diamagnetic currents. Since many
features of the integral quantum Hall effect can be described in
the framework of one particle random magnetic Schr\" odinger
operators it is important to understand their spectral properties
for finite but macroscopic samples with boundaries.

\section{The model}

\indent
The family of random Schr\"odinger operators that we want to study is
\begin{equation}
H_\omega = p_x^2 + \left(p_y-Bx\right)^2 + V_\omega + U_r + U_\ell \qquad \omega \in \Omega
\end{equation}
these are densely defined self-adjoint operators acting in the Hilbert
space $L^2(\mathbb{R}\times\left[-\frac{L}{2},\frac{L}{2}\right],\,\textrm{d} x \,\textrm{d} y)$ with periodic boundary
conditions along $y$: $\psi\left(x,-\frac{L}{2}\right)=\psi\left(x,\frac{L}{2}\right)$,
$x\in \mathbb{R}$. 
$H_0=p_x^2+\left(p_y-Bx\right)^2$ is the kinetic Hamiltonian written in Landau gauge,
its spectrum consists in infinitely degenerate eigenvalues
\begin{equation}
\sigma(H_0)=\sigma_{ess}(H_0)=\left\{(2n+1)B; n\in \mathbb{N}\right\} \; .
\end{equation}
The two confining walls are assumed twice differentiable, strictly monotonic and
satisfy
\begin{eqnarray}
c_1|x+\tfrac{L}{2}|^{m_1}\leq U_\ell(x) \leq
c_2|x+\tfrac{L}{2}|^{m_2} &\quad& \textrm{for  }x\leq
-\tfrac{L}{2} \\
c_1|x-\tfrac{L}{2}|^{m_1}\leq U_r(x) \leq
c_2|x-\tfrac{L}{2}|^{m_2} &\quad& \textrm{for  }x\geq \tfrac{L}{2}
\end{eqnarray}
for some constants $0<c_1<c_2<\infty$ and $2\leq m_1 < m_2<\infty$.
Moreover $U_\ell(x)=0$ for $x\geq -\tfrac{L}{2}$ and $U_r(x)=0$
for $x\leq \tfrac{L}{2}$. We could allow steeper confinements but
the present polynomial conditions turn out to be technically
convenient.
The random potential $V_\omega$ consists of a sum of
local perturbations located at the sites of a finite lattice $\Lambda=\mathbb{Z}^2\cap
\left[X\times \left[-\frac{L}{2},\frac{L}{2}\right]\right]$ where $X$ will be defined
latter. Thus 
\begin{equation}\label{randompot}
V_\omega(x,y)=\sum_{(n,m)\in \Lambda} X_{n,m}(\omega) V(x-n,y-m) \quad \omega
\in \Omega
\end{equation}
where the coupling constants $X_{n,m}$ are i.i.d. random variables
with common bounded probability density $h \in C^2([-1,1])$. The local potential $V$ satisfies $V\in C^2$, $0 \leq V(x,y) \leq V_0< \infty$, $\supp V
\subset \mathbb{B}\left(\boldsymbol{0},\frac{1}{4}\right)$ (the open ball centred at $(0,0)$
of radius $\frac{1}{4}$). $\Omega=[-1,1]^\Lambda$ is the set of all possibles realizations, we will denote by
$\mathbb{P}_\Lambda$ the product measure defined on $\Omega$. 
Clearly for all $\omega \in \Omega$ we have $\|V_\omega\|\leq V_0$. We will
assume that $V_0\ll B$, that is, we work in a strong magnetic field regime or,
equivalently, in a weak disorder regime.\\

Our first result concerns the study of $\sigma(H_\omega)$ in the energy interval
$\Delta_\varepsilon=[B+\varepsilon,B+V_0]$ that lies inside the first Landau
band of the infinite bulk system. In this case the interval $X$, that defines the support of the
random potential along the $x-$direction, is $\left[-\frac{L}{2}+\log
L,\frac{L}{2}-\log L\right]$: we leave a thin strip of size $\log L$ without random
potential along each confining wall.

The second result is about $\sigma(H_\omega)$ inside the first spectral gap of
the infinite bulk system,
more precisely in the energy interval $\Delta=(2B-\delta,2B+\delta)\subset
(B+V_0+\varepsilon,3B-V_0-\varepsilon)$. In this case the random potential fills
the whole space in between the confining walls, that means
$X=\left[-\frac{L}{2},\frac{L}{2}\right]$.\\

Since our system is confined the spectrum is made of discrete eigenvalues.
There exists a natural classification of the eigenvalues via the quantum mechanical current along the periodic
direction. If $\psi$ satisfies the eigenvalue equation $H_\omega\psi=E\psi$ the
current is defined (here) as
\begin{equation}
J_E\equiv (\psi,v_y\psi)
\end{equation}
where $v_y=2(p_y-Bx)$ is the velocity operator in the $y-$direction. Thanks to $J_E$ we can classify the eigenvalues in two classes: the first consists on
those which have $|J_E|>C$ with $C$ a positive constant uniform in $L$, the second
consists on those for which $|J_E|<\epsilon(L)$ with $\epsilon(L)\to 0$ as $L\to
\infty$ (we stress that here $L$ is finite but macroscopic, the limit means that $\epsilon(L)$
is infinitesimally small with $L$). The physical meaning of this classification
is briefly discussed at the end of section 3.\\

The main idea of our approach is to first look at some easier
Hamiltonians and then link them together to get properties on the full
Hamiltonian $H_\omega$. In what follows we do not analyze these easier
Hamiltonians but we just introduce the minimal notations and properties (see
\cite{FM1} and \cite{FM2} for the details).

\section{Main results}

For the analysis of $\sigma(H_\omega)$ in $\Delta_\varepsilon$ we need to know some
properties of the Hamiltonian
\begin{equation}
H_\alpha^0=H_0+U_\alpha \qquad \alpha=\ell,r
\end{equation}
called \emph{pure edge Hamiltonian}. 
Its spectrum is given by
\begin{equation}
\sigma(H_\alpha^0)=\left\{E^\alpha_{n k}; n \in \mathbb{N}, k\in
\tfrac{2\pi}{L}\mathbb{Z}\right\}
\end{equation}

\begin{figure}[!h]
\begin{center}
\input{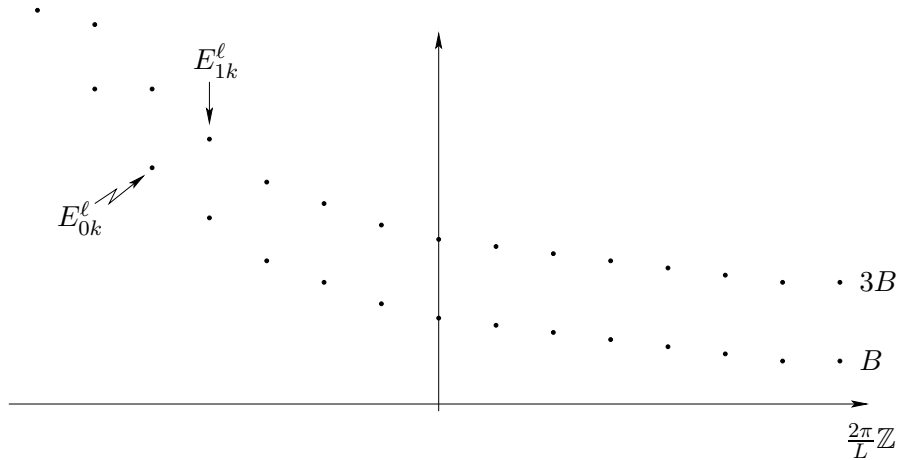}
\end{center}
\caption{\small{\emph{The spectrum of $H_\ell^0$ lies on monotonic decreasing branches.
That of $H^0_r$ lies on similar, but monotonic increasing, branches. The
spectral branches are given by the dispersion relation for $L=\infty$.}}}
\end{figure}

The quantum mechanical currents associated to the eigenfunctions
$\psi_{nk}^\alpha$ whose eigenvalue is in $\Delta_\varepsilon$ (we have $n=0$) satisfy
\begin{equation}
|J_{0k}^\alpha|=|(\psi_{0k}^\alpha,v_y\psi_{0k}^\alpha)|> C
\end{equation}
with $C>0$ a numerical constant independent of $L$ \cite{FM1}. Moreover we will assume the
following
\begin{hyp}\label{hyp1}
Fix $\varepsilon>0$.
There exist $L(\varepsilon)$ and $d(\varepsilon)>0$ such that for
all $L>L(\varepsilon)$
\begin{equation}
\dist \left(\sigma(H_\ell^0)\cap \Delta_\varepsilon,\sigma(H_r^0)\cap
\Delta_\varepsilon\right)\geq \frac{d(\varepsilon)}{L} \; .
\end{equation}
\end{hyp}
\noindent
This hypothesis is important because a minimal amount of
non-degeneracy between the spectra of the two edge systems is
needed in order to control backscattering effects induced by the
random potential. Indeed in a system with two boundaries
backscattering favors localization and has a tendency to destroy
currents.
Remark that this hypothesis can be verified by taking two symmetric confining
potentials $U_\ell(x)=U_r(-x)$ and adding a suitable flux line along the
cylinder axes (see in particular Appendix C in \cite{FM2}).

We also need to know some properties of the \emph{bulk Hamiltonian}
\begin{equation}
H_b= H_0 + V_\omega\; .
\end{equation}
Its essential spectrum is given by the Landau levels and the whole spectrum is
contained in the Landau bands $\sigma(H_b)\subset \bigcup_{n\geq 0} \left[(2n+1)B -
V_0,(2n+1)B + V_0\right]$. We will suppose that the (discrete) spectrum in
$\Delta_\varepsilon$ fullfills the
\begin{hyp}\label{hyp2}
Fix any $\varepsilon > 0$. There exist $\mu(\varepsilon)$ a
strictly positive constant and $L(\varepsilon)$ such that for all
$L>L(\varepsilon)$ one can find a set of realizations of the
random potential $\Omega^{'}$ with
$\mathbb{P}_\Lambda(\Omega^{'}) \geq 1- L^{-\theta}$,
$\theta>0$, with the property that if $\omega\in
\Omega^{'}$ the eigenstates corresponding to $E_\beta^b
\in \sigma(H_b)\cap \Delta_\varepsilon$  satisfy
\begin{equation}\label{H2}
|\psi_\beta^b(x,\bar{y}_\beta)|\leq e^{-\mu(\varepsilon) L}
\qquad,\qquad |\partial_y\psi_\beta^b(x,\bar{y}_\beta)|\leq
e^{-\mu(\varepsilon) L}
\end{equation}
for some $\bar{y}_\beta$ depending on $\omega$ and $L$.
\end{hyp}
\noindent
Since $V_\omega$ is random we expect that eigenfunctions with
energies in $\Delta_\varepsilon$ (not too close to the Landau
levels where the localization length diverges) are exponentially
localized on a scale ${\cal O}(1)$ with respect to $L$.
Inequalities \eqref{H2} are a weaker version of this statement,
and have been checked for the special case where the random
potential is a sum of rank one perturbations \cite{FM3} using the
methods of Aizenman and Molchanov \cite{AM}.
The main consequence of Hypothesis 2 is that a state satisfying \eqref{H2} does not
carry any appreciable current (contrary to the eigenstates of
$H_\alpha^0$) in the sense that 
\begin{equation}
J_\beta^b=(\psi_\beta^b,
v_y\psi_\beta^b)={\cal O}\left(e^{-\mu(\varepsilon) L}\right) \; .
\end{equation}

We are now ready to state our first result on the eigenvalues lying in
$\Delta_\varepsilon=[B+\varepsilon,B+V_0]$.

\begin{thm}
Fix $\varepsilon>0$ and assume that $(H1)$ and $(H2)$ are
fulfilled. Assume $B>4V_0$. Then there exists a numerical constant
$\gamma>0$ and an $\bar{L}\geq L(\varepsilon)$ such that for all $L>\bar{L}$ one can find a set
$\hat{\Omega}\subset \Omega$ of realizations of the random potential $V_\omega$
with $\mathbb{P}_\Lambda(\hat{\Omega})\geq 1- L^{-s}$ $(s\gg 1)$
such that for any
$\omega\in \hat{\Omega}$, $\sigma(H_\omega)\cap
\Delta_\varepsilon$ is the union of three sets $\Sigma_{\ell}\cup
\Sigma_{b}\cup \Sigma_{r}$, each depending on $\omega$ and $L$,
and characterized by the following properties:
\begin{enumerate}
\item[a)] ${\cal E}^\alpha_k\in\Sigma_\alpha$ $(\alpha=\ell,r)$ are a small
perturbation of $E^\alpha_{0k}\in\sigma(H_\alpha^0)\cap
\Delta_\varepsilon$ with
\begin{equation}\label{rt1}
|{\cal E}^\alpha_{k}-E^\alpha_{0k}|\leq  e^{-\gamma B(\log L)^2},
\qquad \qquad \alpha=\ell, r \; .
\end{equation}
\item[b)] For ${\cal E}^\alpha_{k}\in \Sigma_\alpha$ the current ${\cal J}^\alpha_{k}$ of
the associated eigenstate satisfies
\begin{equation}\label{rt2}
\left|{\cal J}^\alpha_{k}-J^\alpha_{0k}\right|\leq e^{-\gamma B(\log L)^2},
\qquad \qquad \alpha=\ell, r \; .
\end{equation}
\item[c)] $\Sigma_b$ contains the same number of energy levels as
$\sigma(H_b)\cap \Delta_\varepsilon$ and $(p\gg1)$
\begin{equation}\label{rt3}
\dist(\Sigma_b,\Sigma_\alpha) \geq L^{-p}, \qquad \qquad
\alpha=\ell, r \; .
\end{equation}
\item[d)] The current associated to each level ${\cal E}_\beta\in \Sigma_b$ satisfies
\begin{equation}\label{rt4}
|{\cal J}_\beta|\leq  e^{-\gamma B(\log L)^2}\; .
\end{equation}
\end{enumerate}
\end{thm}

We now turn to the characterisation of eigenvalues lying in
$\Delta=(2B-\delta,2B+\delta)$. In this case the first easier Hamiltonian is
\begin{equation}\label{edgeh}
H_\alpha=H_0+U_\alpha + V_\omega^\alpha \qquad \alpha=\ell,r
\end{equation}
called \emph{random edge Hamiltonian}. In \eqref{edgeh} the random potential
$V_\omega^\alpha$ is the restriction of $V_\omega$ to $\Lambda_\alpha\subset
\Lambda$, where $\Lambda_\alpha$ is a strip
of size $\sqrt{L}$ in the $x-$direction along the confining walls:
$\Lambda_r= \mathbb{Z}^2 \cap \left[[\frac{L}{2}-\frac{3\sqrt{L}}{4}-1,\frac{L}{2}]\times 
[-\frac{L}{2},\frac{L}{2}]\right]$ and $\Lambda_\ell=\mathbb{Z}^2 \cap
\left[[-\frac{L}{2},-\frac{L}{2}+\frac{3\sqrt{L}}{4}+1]\times[-\frac{L}{2},\frac{L}{2}]
\right]$.
The spectrum of $H_\alpha$ is given by
\begin{equation}
\sigma(H_\alpha)=\left\{E^\alpha_{\kappa}; \kappa \in \mathbb{Z} \right\}
\end{equation}
and $E^\alpha_{\kappa}$ are isolated eigenvalues with accumulation points at the 
Landau levels. 
The quantum mechanical currents $J^\alpha_\kappa$ associated to the energies
in $\Delta$ satisfy
\begin{equation}
|J_{\kappa}^\alpha|=|(\psi_{\kappa}^\alpha,v_y\psi_{\kappa}^\alpha)|> C'
\end{equation}
with $C'>0$ a numerical constant independent of $L$ \cite{FM2}.
By the way remark that, since the random variables in the Anderson potential are i.i.d., $H_\ell$ and $H_r$ are
two independent random Hamiltonians.
Here, as before we suppose that Hypothesis 1 is fulfilled.\\

We now state our second result.

\begin{thm}
Let $V_0$ small enough, fix $\varepsilon>0$ and let
$0<\delta\equiv\delta(V_0)<B-V_0-\varepsilon$. Suppose that $(H1)$
holds. Then there exists $\mu>0$, $\bar{L}\geq L(\varepsilon)$ such that if
$L>\bar{L}$ one can find a set $\hat{\Omega}\subset
\Omega$ of realizations of the random potential $V_\omega$
with $\mathbb{P}_\Lambda(\hat{\Omega})\geq 1-L^{-\nu}$ $(\nu\gg
1)$ such that for all $\omega\in \hat{\Omega}$ the spectrum of
$H_\omega$ in $\Delta=(B-\delta,B+\delta)$ is the unions of two
sets $\Sigma_\ell'$ and $\Sigma_r'$, each depending on $\omega$ and $L$, with the following properties:
\begin{itemize}
\item[a)] ${\cal E}_\kappa^\alpha \in \Sigma_\alpha'$ $(\alpha=\ell,r)$ are a
small perturbation of $E_\kappa^\alpha\in \sigma(H_\alpha)\cap
\Delta$ with
\begin{equation}
|{\cal E}_\kappa^\alpha-E_\kappa^\alpha| \leq
e^{-\mu\sqrt{B}\sqrt{L}} \; .
\end{equation}
\item[b)] For ${\cal E}_\kappa^\alpha \in \Sigma_\alpha'$ the current $\cal{J}_\kappa^\alpha$ of the associated eigenstate satisfies
\begin{equation}
|{\cal J}_\kappa^\alpha-J_\kappa^\alpha| \leq
e^{-\mu\sqrt{B}\sqrt{L}} \; .
\end{equation}
\end{itemize}
\end{thm}

The idea of the proofs of Theorems 1 and 2 is to link the resolvent of the full
Hamiltonian $H_\omega$ to those of the easier Hamiltonians $R_\ell^0(z)$ (resp.
$R_\ell(z)$), $R_r^0(z)$ (resp. $R_r(z)$) and $R_b(z)$. This is achieved via a
decoupling formula for the resolvent \cite{BCD}, \cite{BG}. Using it we can do deterministic estimates on the
norm difference between the projector $P_{H_\omega}(\Gamma)$, associated to $H_\omega$
into the disc with boundary $\Gamma$, and the projector associated to one of the
easier Hamiltonians.
This is done for suitable circles $\Gamma$ in the complex plane and a suitable
set $\hat{\Omega}$ of realizations of the random potential. Using Wegner
estimates on $H_b$ (resp. $H_\alpha$) we control the probability of
$\hat{\Omega}$ and show that it can be made large.\\

Our classification of the spectrum via the quantum mechanical current leads to a
well defined notion of \emph{extended edge states} and \emph{localized bulk
states}. The former are those belonging to $\Sigma_\alpha$ (resp. $\Sigma_\alpha'$),
they are small perturbations of the eigenvalues of $\sigma(H_\alpha^0)$ (resp.
$\sigma(H_\alpha)$) and have a quantum mechanical current of order ${\cal O}(1)$
with respect to
$L$. The latter are those belonging to $\Sigma_b$, and have a infinitesimal current
with respect to $L$ (of order ${\cal O}\left(e^{-\gamma B(\log L)^2}\right)$), they ``arise'' from
the spectrum of $H_b$.
It is interesting to note that our description leads, in the interval inside the
first Landau band, to a spectrum in which extended edge and localized bulk states
are intermixed and in some sense there is no ``mobility edge''.
On the other hand in the interval inside the spectral gap there exists only
extended edge states.

\section*{Acknowledgements} 
C.F. is grateful to the organizers of the conference for the invitation to
report on this work.
The work of C.F. was supported by a grant from the Fonds National Suisse
de la Recherche Scientifique.

\end{document}